\documentclass[sigconf]{acmart}

\AtBeginDocument{%
  \providecommand\BibTeX{{%
    \normalfont B\kern-0.5em{\scshape i\kern-0.25em b}\kern-0.8em\TeX}}}

\usepackage{subfigure} 

\usepackage{bm}
\usepackage{threeparttable}

\copyrightyear{2021} 
\acmYear{2021} 
\setcopyright{acmcopyright}\acmConference[SIGIR '21]{Proceedings of the 44th International ACM SIGIR Conference on Research and Development in Information Retrieval}{July 11--15, 2021}{Virtual Event, Canada}
\acmBooktitle{Proceedings of the 44th International ACM SIGIR Conference on Research and Development in Information Retrieval (SIGIR '21), July 11--15, 2021, Virtual Event, Canada}
\acmPrice{15.00}
\acmDOI{10.1145/3404835.3462884}
\acmISBN{978-1-4503-8037-9/21/07}

\AtBeginDocument{%
  \providecommand\BibTeX{{%
    \normalfont B\kern-0.5em{\scshape i\kern-0.25em b}\kern-0.8em\TeX}}}

\begin{document}

\fancyhead{}
\title{ One Person, One Model, One World: Learning Continual User Representation without Forgetting
}

\author{Fajie Yuan$^{\dagger,\dotplus}$, Guoxiao Zhang$^{\dotplus}$, Alexandros Karatzoglou$^{\ddagger}$, Joemon Jose$^{\wr}$, Beibei Kong$^{\dotplus}$,   Yudong Li$^{\dotplus}$}

\affiliation{\institution{$^{\dagger}$Westlake University, China \qquad $^{\dotplus}$Tencent, China, \qquad $^{\ddagger}$Google, UK \qquad $^{\wr}$University of Glasgow, UK}}




\email{yuanfajie@westlake.edu.cn, suranzhang@tencent.com, alexandros.karatzoglou@gmail.com}


 \thanks{$^{\dagger,\dotplus}$This work was done when Fajie worked at Tencent (past affiliation) and Westlake University (current affiliation)}

	\begin{abstract}

Learning user representations is a vital technique toward  effective user modeling and personalized recommender systems.
Existing approaches often derive an individual set of model parameters for each task by training on separate data. However, the representation of the same user potentially has some commonalities, such as preference and personality, even in different tasks. As such, these separately trained representations could be suboptimal in performance as well as inefficient in terms of parameter sharing.

In this paper, we delve on research to continually learn user representations task by task,  whereby new tasks are learned while using partial parameters from old ones. A new problem arises since when new tasks are trained,  previously learned parameters are very likely to be modified, and as a result,  an artificial neural network (ANN)-based model may lose its capacity to serve for  well-trained previous tasks forever, this issue is termed catastrophic forgetting. To address this issue, we present  \emph{Conure} the first \underline{con}tinual, or lifelong, \underline{u}ser  \underline{re}presentation learner ---  i.e., learning new tasks over time without forgetting old ones. Specifically, we propose iteratively removing less important weights 
of old tasks in a deep user representation model, motivated by the fact that neural network models are 
usually over-parameterized. In this way, we could learn many tasks with a single model by reusing the important weights, and modifying the less important weights to adapt to new tasks.  We conduct extensive experiments on two real-world datasets with nine tasks and show that \emph{Conure} largely exceeds the
standard model that does not purposely preserve such old ``knowledge'', and 
performs competitively or sometimes better than models which are  trained either individually for each task or simultaneously by merging all task data. 
\end{abstract}


\maketitle

\section{Introduction}
\label{Introduction}

In the last decade,  social medial  and e-commerce systems, such as TikTok, Facebook and Amazon, have become increasingly popular and gained success due to the convenience they provide in people's lives. For example, as the biggest social network, Facebook has over 2.6 billion monthly active users.\footnote{\scriptsize\url { https://www.statista.com/statistics/264810/number-of-monthly-active-facebook-users-worldwide}} On the other hand, a large number of user behavior feedback (e.g., clicks, likes, comments and shares) is created every day on these systems. An impressive example is TikTok, where users can easily watch hundreds of short videos per day 
given that the play duration per video takes usually less than 30 seconds~\cite{yuan2020parameter}.  

 A large body of research~\cite{guo2017deepfm,yuan2016lambdafm,zhou2020s3,hidasi2015session,yuan2019simple,chen2021user,wang2020stackrec,sun2020generic} has demonstrated that the user behavior signals can be used to model their preference so as to provide personalized services, e.g., for recommender systems. However, most of these work focuses only on the tasks of user modeling (UM) or item recommendation on the same platform, from where the data comes. Unlike these  works, recently~\cite{yuan2020parameter} took an important step, which revealed that the user representations learned from an upstream  recommendation task could be 
a generic representation of the user and could be directly transferred to improve a variety of dissimilar downstream tasks. To this end, they proposed a two-stage transfer learning paradigm, termed PeterRec, which first performs self-supervised pretraining on user behavior sequences,
and then performs task-specific supervised finetuning on the corresponding downstream tasks.



Despite that PeterRec has achieved some positive transfer, the  downstream tasks it served for, however, are trained individually. These tasks may share substantial similarities
 in practice if the same users are involved. E.g., users who retweet a message posted on Twitter tend to give it a thumb-up as well. That is, the task of thumb-up prediction  has some correlations with the task of retweet prediction. Arguably, we believe  learning  user representations from many tasks is important and could potentially obtain better performance on related  tasks.
Besides, training tasks individually
 requires additional storage overhead to keep parameters per model per task; otherwise, training them one by one with a single model may lead to catastrophic forgetting~\cite{kirkpatrick2017overcoming,mallya2018piggyback}.
In parallel, another line of  work usually
perform multi-task learning (MTL) on related tasks ~\cite{ni2018perceive,ma2018modeling}. This could be beneficial as well but sometimes infeasible 
since training data for all tasks might not always be simultaneously available. Moreover, some incoming tasks may not have overlapping users, making such joint learning-based methods often infeasible.
\begin{figure}[t]
	\centering
	\small	
	\includegraphics[width=0.45\textwidth]{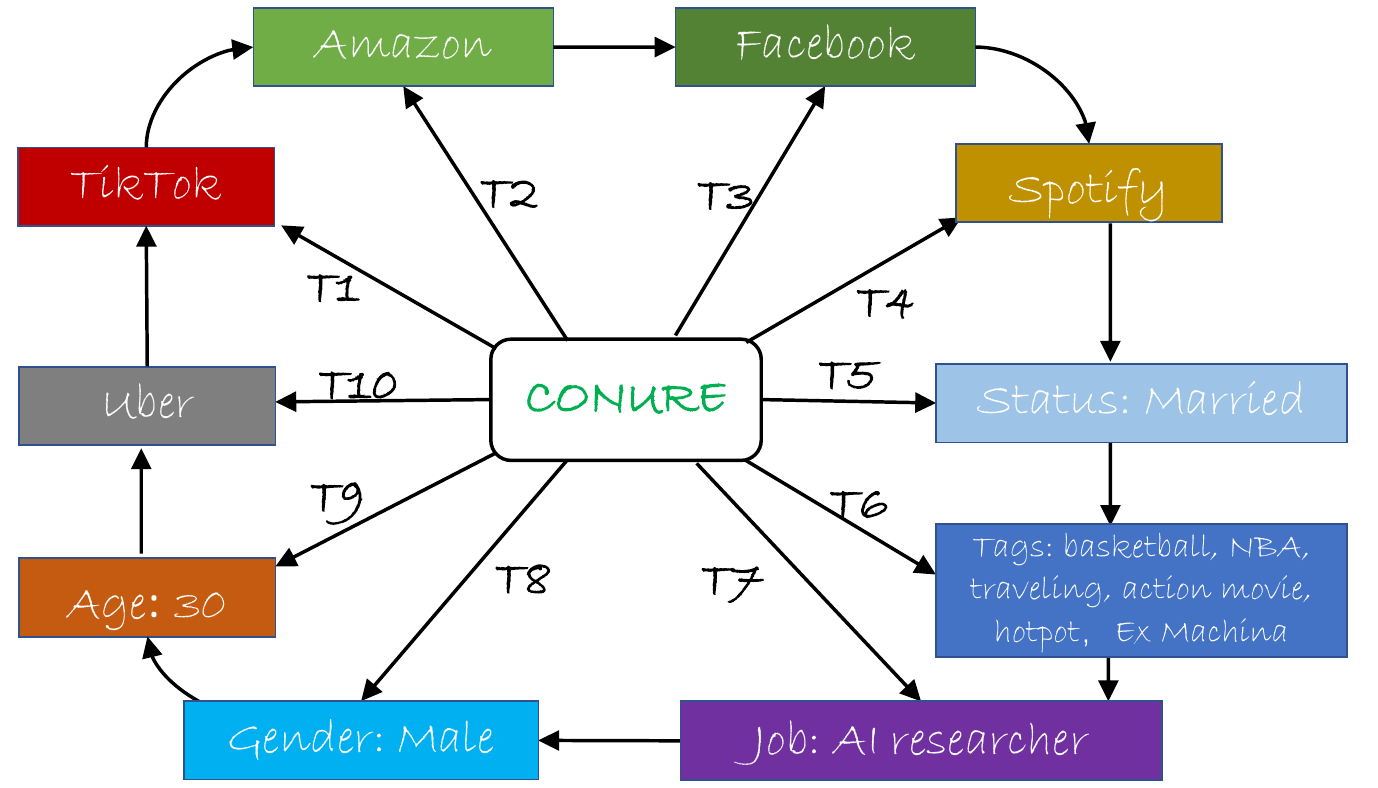}	
	\caption{\small Illustration of one person, one model, one world. `T' is short for task (e.g., item recommendation (i.e., T1, T2, T3, T4, T10) or profile prediction (i.e., T5, T6, T7, T8, T9)). While a person played different roles in the 10 scenarios in his (real or virtual) world,   a lifelong UR model, for example, \emph{Conure}, should be able to learn user representations continually task by task and then serve all of them after a round of training.
	}
	\label{opomow}
		\vspace{-0.05in}
\end{figure}

To deal with the above-mentioned issues, we  explore a promising but more challenging learning paradigm for user modeling --- i.e., lifelong user representation (UR) learning over tasks. Our goal is to develop an artificial neural network (ANN)-based UR model that not only provides universal user representations, but also has the continuous learning ability throughout its lifespan:
quickly learning new abilities based on previously acquired knowledge, and being immune to forgetting old knowledge.
Moreover, 
the proposed UR model should build up and modify representations for \underline{each person} with only \underline{one backbone} network architecture, whereby all roles the person played in their \underline{individual} (real or virtual) \underline{world}  can be well described. Ideally, with such a comprehensive UR model,  
 our understanding towards user needs, preferences, cognitive and behavioural characteristics  could enter a new stage.
We refer to this goal as One Person, One Model, One World, as shown in Figure~\ref{opomow}. 

To motivate this work, we first perform ablation studies to show two unexplored phenomena for deep UR models:  i) sequentially
learning different tasks and updating parameters for a single ANN-based UR model leads to  catastrophic forgetting, and correspondingly, the UR model loses its prediction ability for old tasks that were trained before; ii) removing a certain percentage of  unimportant/redundant
  parameters for a well-trained deep UR model does not cause irreversible degradation on its prediction accuracy. Taking inspiration from the two insights, we propose a novel \underline{con}tinual, or lifelong, \underline{u}ser  \underline{re}presentation learning framework, dubbed as \emph{Conure}. \emph{Conure}  is endowed the lifelong learning capacity for a number of tasks related to user profile prediction and item recommendation, where it addresses the forgetting issue for old tasks by 
 important knowledge
retention, and learns
 new tasks by
 exploiting parameter redundancy.
We summarize our main contributions as follows.
\begin{itemize}
\item  We open a new research topic
and
formulate the first UR learning paradigm that deals with a series of  \textbf{\textit{different}} tasks coming either sequentially or separately.\footnote{\scriptsize Note recent work in~\cite{mi2020ader,ren2019lifelong} also introduced a `lifelong' learning solution for RS, the main difference between  our paper and them is described in Section~\ref{continuallearn}. }
Besides,
we show in-depth empirical analysis for the forgetting problem  and network redundancy in deep UR models under the proposed lifelong learning setting. 
\item We present \emph{Conure}, which  could compact multiple (e.g., 6) tasks sequentially into a single deep UR model without network expansion and forgetting. \emph{Conure} is conceptually simple, easy to implement, and applicable to a broad class of sequential encoder networks.
\item  We instantiate  \emph{Conure} by using temporal convolutional network (TCN)~\cite{yuan2019simple} as  the backbone network for case study, and report important results for both TCN and the self-attention based network (i.e., Transformer~\cite{vaswani2017attention}).
\item We provide many useful insights regarding performance of various learning paradigms in the field of recommender systems and user modeling. We demonstrate that \emph{Conure}  largely exceeds
its counterpart that performs the same continual learning process but without purposely preserving old knowledge. Moreover, \emph{Conure}   matches or exceeds separately trained models, typical transfer learning and multi-task learning approaches, which require either more model parameters or more training examples.
\end{itemize}

\section{Related Work}
Our work intersects with research on user modeling (UM) and recommender systems (RS), transfer learning (TL),
 and continual learning (CL). We briefly review recent advances below.

\begin{figure*}[t]
	\centering
	\small	
	\includegraphics[width=0.99\textwidth]{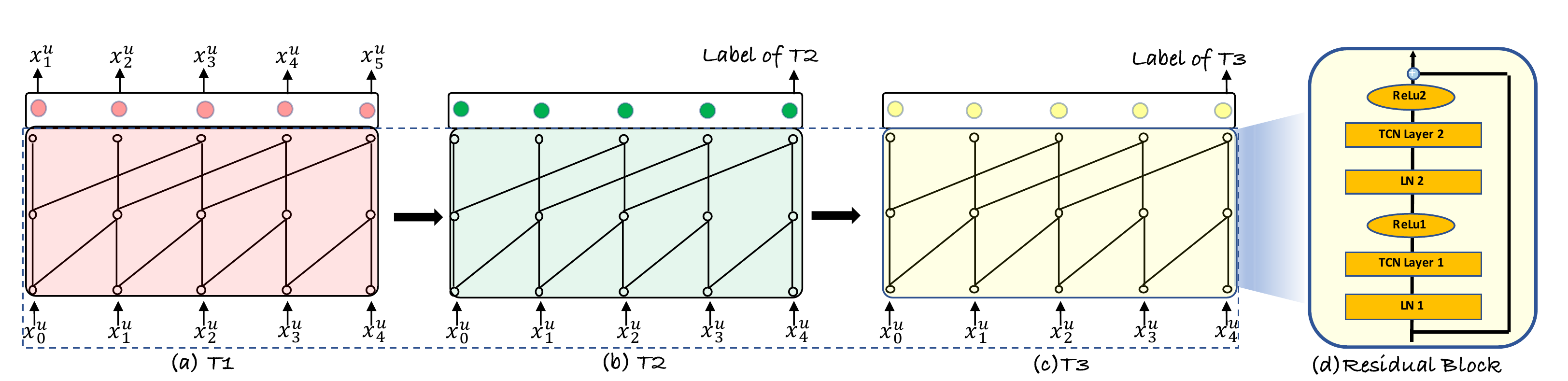}	
	\caption{\small The continual learning framework with the TCN architecture without purposely preserving prior knowledge.  
	(a)(b) and (c) show the network architectures of $T_1$, $T_2$,  and $T_3$, respectively.   The dashed frame denotes the backbone  (i.e., the so-called user representation) encoder network, whose parameters are shared and will be modified (from red in $T_1$, green in $T_2$, to yellow in $T_3$) during training of each task. The top layers with red, green and yellow colors on (a) (b) and (c) respectively are task-specific prediction layers.
	(4) is a typical residual block of TCN.  }
	\label{prelim}
\end{figure*}

\subsection{User Modeling and Recommendation}
User modeling refers to the process of obtaining the user profile, 
which is a conceptual understanding of the user. 
It is an important step towards personalized recommender systems. One common research line of UM  is based on representation learning, where users or their behaviors are  modeled and represented by certain types of machine learning algorithms~\cite{ni2018perceive,yuan2019simple,wang2020stackrec,zhou2020s3}. These well-trained digital user models are often called user representations.

Over recent years, deep neural networks have become dominant techniques for user representation learning. Among many, deep structured semantic models (DSSM)~\cite{huang2013learning}, deep or neural factorization machines (DeepFM/NFM) ~\cite{guo2017deepfm,hexiagnan2017neuralf} have become some  representative work based on supervised representation learning. However,  these learned UR models have been shown useful only for a specific task.
One reason is that 
the supervised  learning objective  functions usually  focus  upon a specific goal only~\cite{yuan2020parameter}, which  may not generalize well to other tasks. 


Differently, PeterRec presented a self-supervised pretraining approach based on the sequential recommendation model NextItNet~\cite{yuan2019simple}.
The pretraining process used is the prediction of the next user-item interaction in the user behavior sequence.
By modeling the inherent relations of behavior sequences, the learned user representations become universal rather than specialized, and thereby can be used for many other tasks.
Nevertheless, PeterRec  enables only one-time TL between the first task (say, $T_1$) and the other task (e.g, $T_2$ or $T_3$), i.e., $T_1\rightarrow {T_2}, \  T_1\rightarrow T_3,\ ..., \ T_1\rightarrow T_N$
rather than the continual TL among all tasks, e.g., $T_1\rightarrow {T_2} \rightarrow T_3, ... ,\rightarrow T_N$ or  $T_1\rightarrow {T_N} \rightarrow T_3, ... ,\rightarrow T_2$.\footnote{\scriptsize `$\rightarrow$' denotes the direction of TL.}
In this paper, we design and instantiate  \emph{Conure} based on PeterRec-style  network architecture, which can be seen as an extension of PeterRec towards continual UR learning.


\subsection{Transfer Learning}
TL is typically based on a two-stage training paradigm: first pretraining a base model on the source dataset and then finetuning a new model on the target dataset with part  or all of the pretrained parameters as initialization. Following PeterRec, we choose the well-known temporal (a.k.a. dilated) convolutional network (TCN)~\cite{yuan2019simple,yuan2020future}  as the pretrained base model for case study  given its linear complexity and superb performance in modeling sequences~\cite{bai2018empirical,wang2019towards,tanjim2020dynamicrec,wang2020stackrec,zhao2020amer,oord2016wavenet,chen2021user}.
 \emph{Conure} is more related to the finetuning stage, which can be in general classified into the four types~\cite{devlin2018bert,yuan2020parameter}: i) finetuning only the softmax layer with the pretrained network as a feature extractor; ii) finetuning some higher layers while keeping the bottom layers frozen; iii) finetuning the entire pretrained model; and iv) finetuning only some newly added adaptor networks like PeterRec. 



 \subsection{Continual Learning }
 \label{continuallearn}
 CL refers to the continuous learning ability of an AI algorithm throughout its lifespan.  
%
  It is regarded as an important step towards general machine intelligence~\cite{parisi2019continual}. While it has been  explored in computer vision (CV)~\cite{li2017learning,mallya2018packnet,zenke2017continual,golkar2019continual,wang2020learn} and robot learning~\cite{thrun1995lifelong,mitchell2018never}, to our best knowledge, such task-level CL has never been  studied for user modeling and recommender systems. In fact, it is largely unknown whether the learning paradigms, frameworks, and methodologies 
 for other domains are useful or not to address our problem. 
  Meanwhile, there are also some  recent work in ~\cite{ren2019lifelong,qi2020search,mi2020ader}  claiming that RS models should  have the so-called `lifelong' learning capacity. However, their methodologies are  designed only to  model long-term user behaviors or new training data from the same  distribution or task, which distinguish from \emph{Conure}, capable of sequentially or separately learning very \textbf{\textit{different}} tasks. 

\section{PRELIMINARIES}
 We  begin with formulating the continual learning (CL) paradigm for user representations.  Then, we perform experiments to verify the impacts of  the catastrophic forgetting and
the over-parameterization issues for deep user representation models.


\subsection{Task Formulation}
Suppose we are given a set of consecutive tasks $\textbf{T}=\{T_1, T_2,...,  T_N\}$,
 where $\textbf{T}$ is theoretically unbounded and  allowed to increase new tasks throughout the lifespan of a CL algorithm. First we need to learn the base representations for users in $T_1$, and then ensure the continual learning of them if they appear in the following tasks, i.e., $\{T_2,\ ..., \ T_N\}$, so as to achieve more comprehensive representations.
Denote $\mathcal{U}$ (of size $|\mathcal{U}|$) as the set of users in $T_1$.
Each instance in $T_1$ contains a userID  $u \in \mathcal{U}$, and his/her interaction sequence  $\bm{\mathrm{x}}^u=\{x_0^u, ..., x_{n}^u \}$ ($x_i^u \in \mathcal{X}$), 
i.e., $(u,\bm{\mathrm{x}}^u) \in  T_1$, 
where $x_t^u$ is the $t$-th interaction  of $u$ and  $ \mathcal{X}$  (of size $|\mathcal{X}|$)  is the set of items. For example, $T_1$ can be  a video recommendation task where a number of user-video watching interactions are often available. Note that since in $T_1$ we are learning the base user representations, we assume that users in  $T_1$ have  at least several interactions for learning, although  theoretically \emph{Conure}  works even with one interaction. 
On the other hand, each instance in $\{T_2,..., T_N\}$ is formed of a userID $u \in \tilde{\mathcal{U}}\subseteq \mathcal{U} $ and a supervised label $y \in \mathcal{Y}$ (of size $|\mathcal{Y}|$), i.e., $(u,y) \in T_i$.  If $u$  has more than one label, say $g$, then there will be $g$ instances for $u$.
In our CL setting, $\{T_2,..., T_N\}$ can be different tasks, including various profile (e.g., gender) prediction and item recommendation tasks, where $y$ denotes a specific class (e.g., male or female) or an itemID, respectively. After training of $\textbf{T}$, our \emph{Conure} should be able to serve all tasks in $\textbf{T}$ by one individual model.


\subsection{ Learning Sequential Tasks with TCN}
\label{lst}

In the training stage, \emph{Conure}  learns tasks in $\textbf{T}$ one by one (e.g., $T_1\rightarrow {T_2} \rightarrow T_3, ... ,\rightarrow T_N$) with only one backbone network, as shown in Figure~\ref{prelim}.
We present this vanilla CL procedure as follows.

\noindent{\textbf{Training of $T_1$}}: 
As the first task, we should learn the base user representation (UR) which is expected to be  universal rather than task-specific. To do so, we model the  user interaction sequence $\bm{\mathrm{x}}^u$ by an autoregressive (a.k.a. self-supervised) learning manner.
Such training method was introduced into sequential recommender systems by NextItNet, which is also 
very popular in computer vision (CV)~\cite{van2016pixel}, natural language processing (NLP)~\cite{vaswani2017attention,geng2021iterative}. Formally,
the joint distribution of a user sequence  is represented as  the product of conditional distributions over all user-item interactions:
\begin{equation}
\label{l2r}
p(\bm{\mathrm{x}}^u;\Theta)=\prod_{j=1}^{n}p(x_j^u|x_0^u, .., x_{j-1}^u;\Theta)
\end{equation}
where the value $p(x_j^u|x_0^u, ..., x_{j-1}^u)$ is the probability of the $j$-th interaction $x_j^u$ conditioned on all its past interactions $\{x_0^u,...,x_{j-1}^u\}$. Figure~\ref{prelim} (a) illustrates this  conditioning scheme with TCN as the backbone network (described later).
After training of $T_1$,  the backbone (i.e., the so-called user representation model) could be transferred for many other tasks $T_i$ ($i\geq 2$)  according to the study in ~\cite{yuan2020parameter}.

\noindent{\textbf{Training of $T_i$}}: The training of $T_i$  ($i\geq 2$) is shown in  Figure~\ref{prelim} (b) and (c). $T_i$ is connected with  $T_1$ by userID.
For each instance $(u,y)$ on $T_i$, 
we take the interaction sequence of $u$ (in $T_1$) as input and feed it to its sequential encoder
network, i.e.,
the backbone of $T_1$ as well.
Let $\bm{E}_0 \in  \mathbb{R}^{n \times f}$ be the embedding matrix of  $\bm{\mathrm{x}}^u$, where $f$ is the embedding size. After passing it through the encoder network,
%
we obtain the final hidden layer, denoted as  $\bm{E} \in  \mathbb{R}^{n \times f} $. Then, a dense prediction  (or softmax) layer is placed on the last index vector of  $\bm{E}$, denoted by $\bm{g}_{n-1} \in  \mathbb{R}^{f}$. 
Finally, we can predict scores $\bm{h} \in \mathbb{R}^{|\mathcal{Y}|}$ with respect to all labels in $\mathcal{Y}$ by $\bm{h} =\bm{g}_{n-1}\bm{W}+\bm{b}$, where  $\bm{W} \in  \mathbb{R}^{f \times |\mathcal{Y}|} $ and $\bm{b} \in \mathbb{R}^{|\mathcal{Y}|} $ denote the projection matrix and bias term, respectively.


In terms of the training loss of $T_i$, one can apply either a ranking or  a classification  loss.   In this paper, 
we report results using BPR~\cite{rendle2009bpr} loss
with the popular item-frequency\footnote{\scriptsize Item-frequency based negative sampler has shown better performance than the random sampler in much literature w.r.t. the top-N metrics, such as MRR@N and NDCG@N~\cite{hidasi2015session}} based negative sampling (see~\cite{yuan2016lambdafm})
for top-N item recommendation tasks and
the cross-entropy classification loss for profile prediction tasks.

 \noindent{\textbf{Backbone Network}}:
  For better illustration, we instantiate \emph{Conure} using the TCN  architecture in the following despite that the  framework is network-agnostic. Apart from the embedding layer,
 the TCN encoder is  composed  of  a  stack  of  temporal  convolutional layers,  every two of which  are wrapped  by  a  residual  block  structure, as shown in Figure~\ref{prelim} (d).  The $l$-th residual block is formalized as
 \begin{equation}
\bm{E}_{l} = \bm{F}_{l}(\bm{E}_{l-1}) + \bm{E}_{l-1}
\end{equation}
where $\bm{E}_{l-1}$ and $\bm{E}_{l}$ are the input and output of the $l$-th residual block respectively,
and $\bm{F}$ is the residual mapping to be learned
 \begin{equation}
\bm{F}_{l}(\bm{E}_{l-1})=  \sigma( \phi_2( {LN}_2(\sigma (\phi_1 ({LN}_1 (\bm{E}_{l-1}))))))
\end{equation}
where $\sigma$ is the ReLu~\cite{nair2010rectified} operation, $LN$ is layer normalization~\cite{ba2016layer} and $\phi$ is the TCN layer. Biases are omitted for simplifying notations.
\subsection{ Forgetting from $T_1$ to $T_2$}
\label{fogetting}
\begin{figure}
	\small
		\vspace{-0.1in}
	\centering     
	\subfigure[\scriptsize Only training $T_1$ ]{\label{yahoo-alpha}\includegraphics[width=0.235\textwidth]{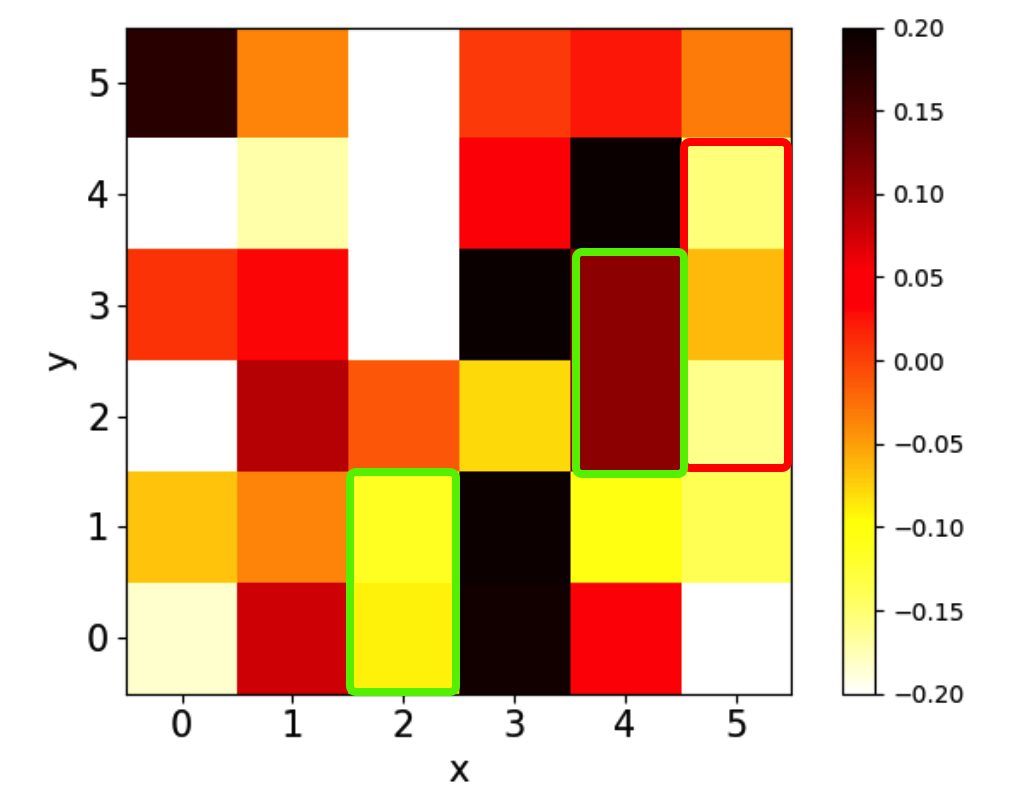}}
	\subfigure[\scriptsize After Training $T_2$  ]{\label{yahoo-alpha}\includegraphics[width=0.235\textwidth]{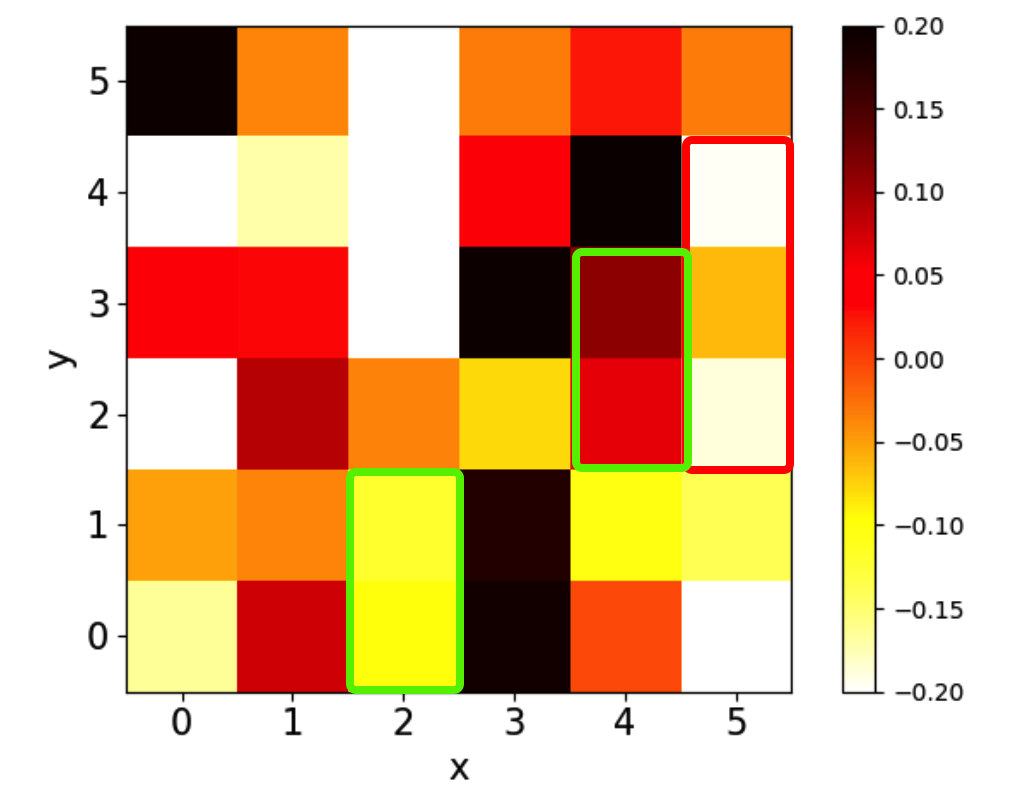}}
	\subfigure[\scriptsize Only training $T_1$ ]{\label{yahoo-alphazero}\includegraphics[width=0.235\textwidth]{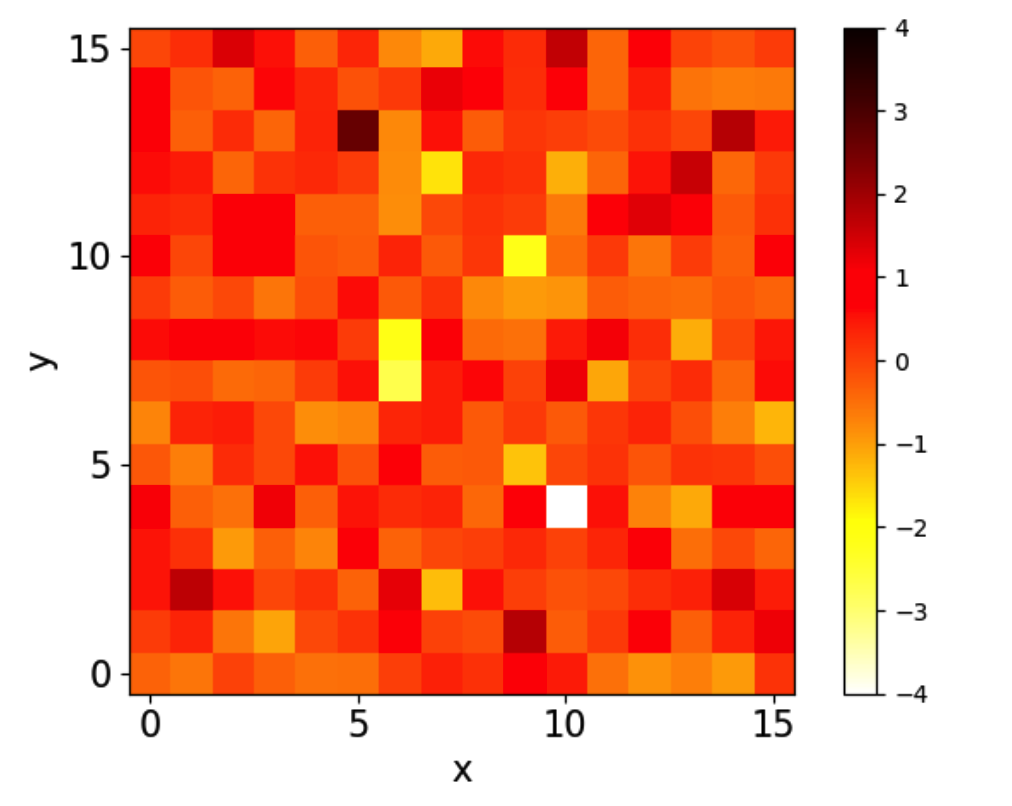}}
	\subfigure[\scriptsize After training $T_2$  ]{\label{yahoo-alpha}\includegraphics[width=0.235\textwidth]{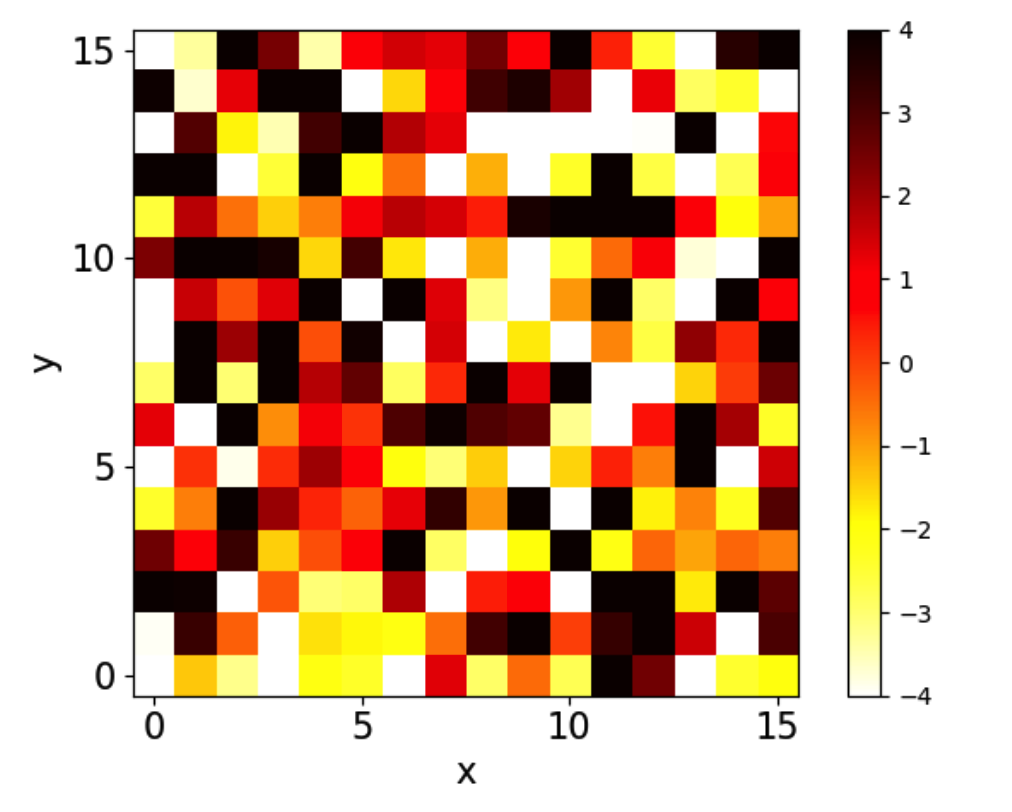}}
	\caption{\small Forgetting issue during continual learning. (a) and (b) represent a reshaped 2-D (i.e., from $1\times3\times12$ to $6\times6$)  convolution kernel of the last hidden layer, while (c) and (d) represent a reshaped (i.e., from $256$ to $16\times 16$) 2-D  matrix
		of $\bm{g}_{n-1}$.  Significantly different pixels on (a)  (b) are marked by the red \& green frames. }.
	\label{Cataforget}
		\vspace{-0.1in}
\end{figure}

We investigate the catastrophic forgetting issue by sequentially learning $T_1$ and $ T_2$.  Since the model on $T_2$ shares the same backbone network as $T_1$,  the optimization of it for $T_2$ will  also lead to the parameter modification of  $T_1$. We show the comparisons of weights and the final hidden vector $\bm{g}_{n-1}$ (of a randomly selected user-item interaction sequence) before and after the training of $T_2$ on Figure~\ref{Cataforget}. Note  in our experiments we use the convolution kernel $1\times3\times256$, where $1\times3$ and 256 are the kernel size and number of channels, respectively. To clearly demonstrate the difference, we select channels from the top $12$-th indices, i.e., $1\times3\times12$ and depict them on (a) and (b).

 At first glance, (a) and (b)
 look quite similar. It seems that the forgetting issue is not as serious as we imagined. However, we  notice that the prediction accuracy (mean reciprocal rank MRR@5) drops drastically from $0.0473$  to $0.0010$ (almost completely forgetting) when performing exactly the same evaluation on $T_1$ with the newly optimized parameters by training $T_2$. The degradation of performance  implies more serious forgetting problem, compared with deep models in other domains~\cite{kirkpatrick2017overcoming,li2017learning}.
 To  identify the cause, we further check the changes of $\bm{g}_{n-1}$, which directly determines the final results  together with the prediction layer of $T_1$. Clearly, the subfigure on (c) and (d) shows very big changes of $\bm{g}_{n-1}$ after learning $T_2$. In fact,  we find that most of the weights on (a) have already been modified, but with a relatively small range (around $\pm 20\%$), which is thus not very  visible on figures. For instance, the first value with $(x,y) =(0,0)$ changed from $-0.1840$ on (a) to $-0.1645$ on (b). It is reasonable that such small weight changes on many layers may incur cumulative effect, and lead to largely different outputs. i.e., the so-called catastrophic forgetting. 

\subsection{ Over-parameterization phenomenon}
\begin{figure}
	\small
		\vspace{-0.1in}
	\centering     
	\subfigure[\scriptsize $f=256$ with 16 CNN layers ]{\label{yahoo-alphazero}\includegraphics[width=0.23\textwidth]{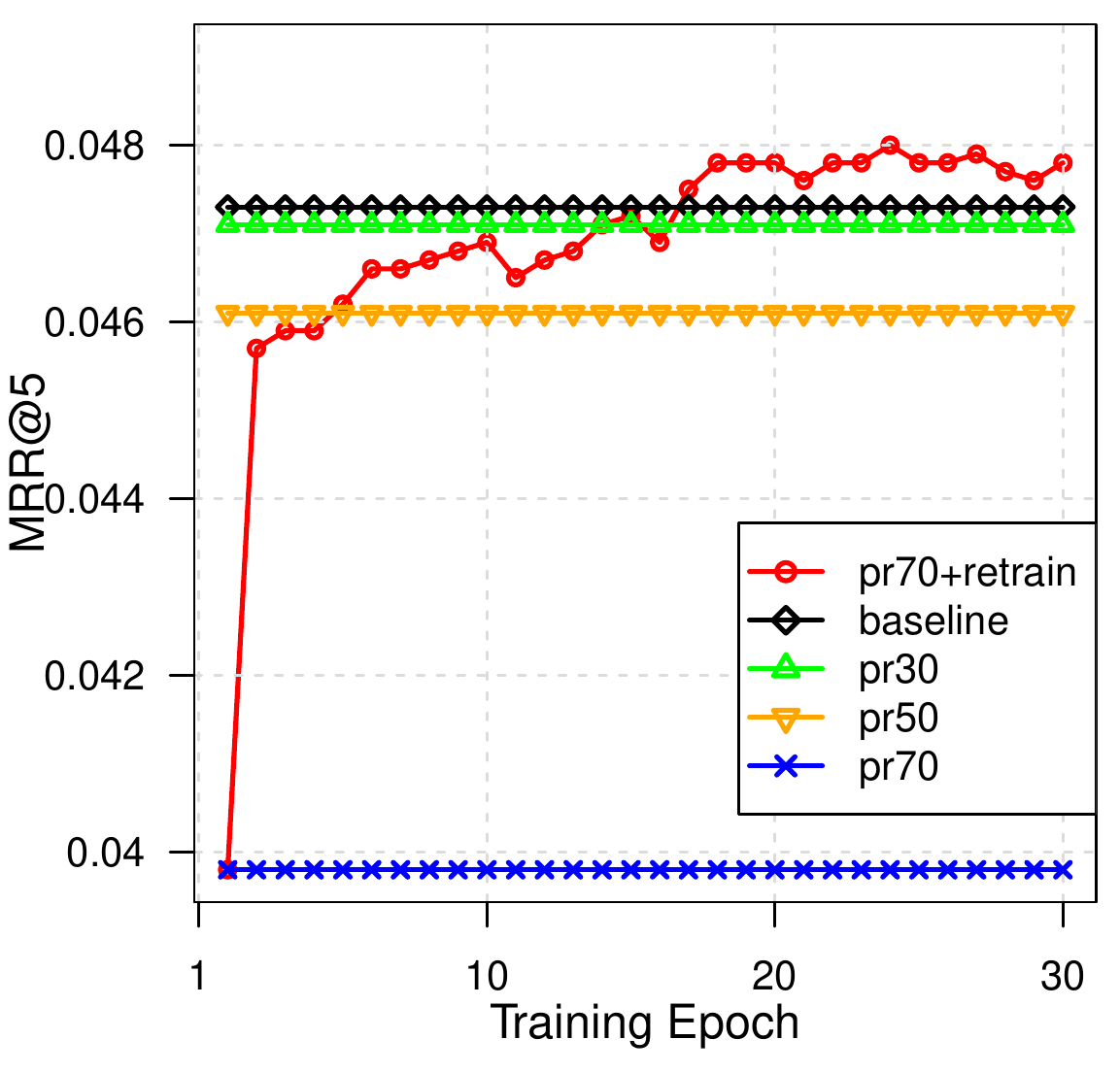}}
	\subfigure[\scriptsize  $f=64$  with 8 CNN layers ]{\label{yahoo-alphazero}\includegraphics[width=0.23\textwidth]{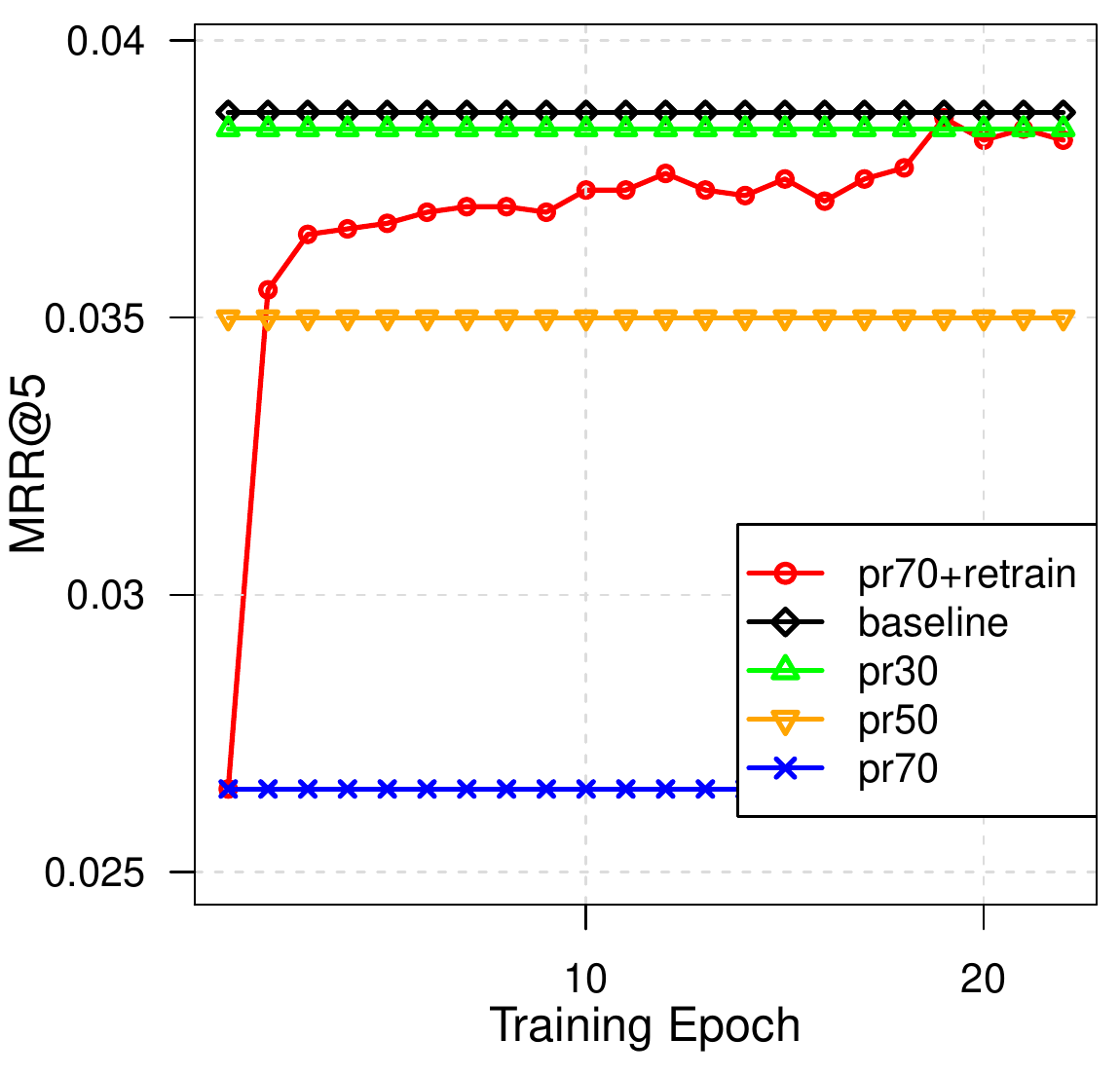}}
	\caption{\small Network trimming.  ``pr30", ``pr50" \& ``pr70"  denote pruning  30\%, 50\%, 70\% parameters of the convolutional layers, respectively. ``pr70+retrain" means performing retraining on the pruned network. Each epoch has $1000*32$ examples, where 32 is batch size.  } 
	\label{pruning}
\end{figure}
We remove a certain percentage of unimportant/redundant  parameters for each  TCN layer trained after $T_1$. 
The importance of a parameter is  measured by its absolute value sorted in the same layer. 
This process is often referred to as network trimming or pruning~\cite{hu2016network}, which was originally invented for model compression~\cite{han2015deep,meng2020filter}. 
We report the pruning results in Figure~\ref{pruning}.  It shows that simply removing unimportant parameters results in a loss in accuracy --- the more are pruned, the worse it performs. E.g, pruning 70\% parameters hurts the accuracy seriously due to the sudden change in network connectivity. 
Fortunately, performing retraining on the pruned network (i.e., ``pr70+retrain'') regains its original accuracy quickly, as shown on both (a) \& (b). This, for the first time, evidences that over-parameterization or redundancy widely exists in the deep user representation model. Moreover, we note that even the network with a much smaller parameter  size,  i.e., having not reached its full ability,
is still highly redundant, as shown on~(b). 

\section{Conure}
\begin{figure}[t]
	\centering
	\small	
	\includegraphics[width=0.485\textwidth]{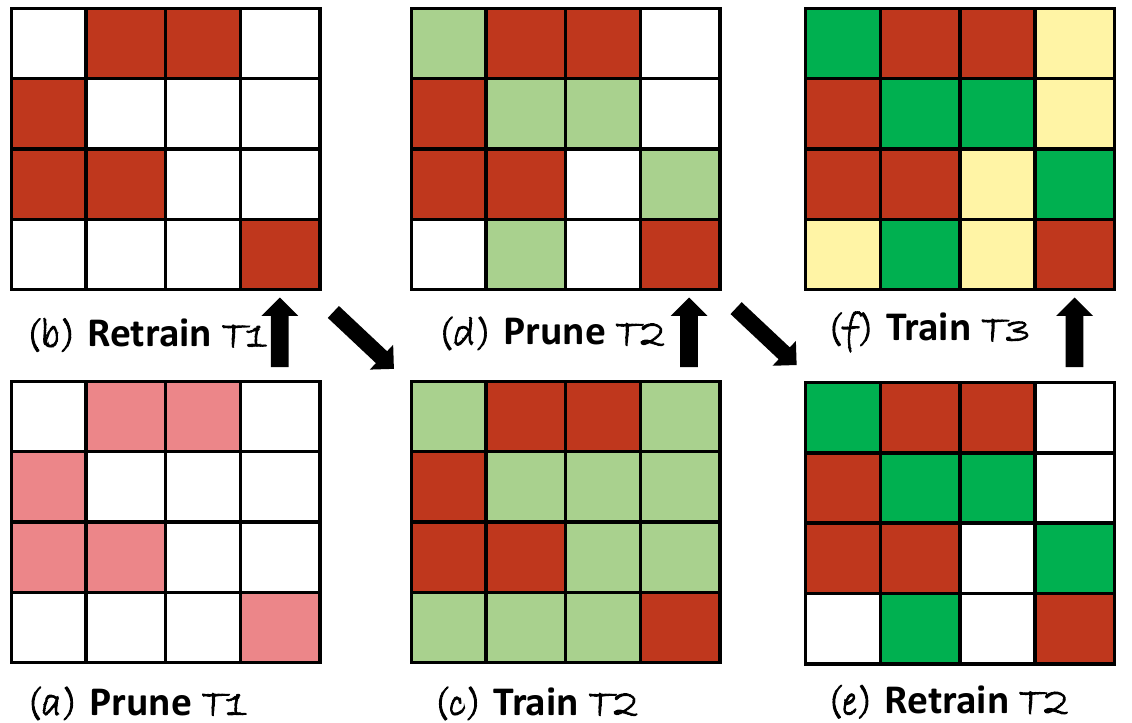}	
	\caption{\small
	Overview of  \emph{Conure}.
		The  subfigures from (a) to (f) denote  parameter matrices of  $T_1$, $T_2$,  and $T_3$. The black arrows denotes the continual learning process. The white cells represent invalid parameters (i.e. set them to zero) for the current task, while the red (including both light red and dark red), green and yellow cells represent task-specific parameters for $T_1$, $T_2$, and $T_3$, respectively. Parameters of $T_2$ include both red and green cells, similarly, parameters of  $T_3$ include all red, green and yellow cells. }
	\label{conure_para}
		\vspace{-0.1in}
\end{figure}

Driven by the above insights, we could develop \emph{Conure} for multiple tasks by exploiting parameter redundancy in  deep user representation models:
first removing  unimportant parameters to free up space for the current task,  then learning new tasks and filling  task-specific parameters into the freed up capacity. 
To obtain positive transfer learning,  important parameters from past tasks should be employed; to prevent forgetting, these  important parameters should be kept fixed when learning new tasks. 
Figure~\ref{conure_para} gives an overview of  \emph{Conure}. 

Specifically, we begin by assuming that the base user representations have been  obtained by training $T_1$ (see Figure~\ref{prelim} (a)). Before learning a new task (e.g., $T_2$), we first perform  network pruning~\cite{han2015deep} to remain only a percentage of important parameters (light red cells in Figure~\ref{conure_para} (a)) on the backbone network. After pruning, the model performance
could be affected because of the big changes in network structure.
We perform retraining (Figure~\ref{conure_para} (b)) over these important parameters  on the pruned architecture   so as to regain its original performance, 
After
this step, there are some free parameters left (the white cells in (b)), which are allowed to be optimized when  learning a new task.
In this way, when a new task arrives, \emph{Conure} keeps learning it by only back propagating these free parameters while the remaining important parameters are hereafter kept fixed (Figure~\ref{conure_para} (c)) for all future tasks. Next, by iteratively performing such network trimming (Figure~\ref{conure_para} (d)) \& retraining (Figure~\ref{conure_para} (e)) on  parameters of the current task, the user model could accommodate more tasks. Our idea here is initially inspired by ~\cite{meng2020filter} to some extent. 
The key difference is that  unimportant/redundant  parameters in~\cite{meng2020filter} are replaced into important parameters from an external network for accuracy improvement, while unimportant  parameters in \emph{Conure} are re-optimized based on the new task so as to realize continual representation learning. We notice that similar attempts have been recently proven effective for solving problems in other research fields~\cite{wang2020learn,geng2021iterative}.
In what follows, we provide a  detailed 
explanation by specifying the TCN recommender as the backbone.




\subsection{Methodology Details}


\noindent\textbf{Redundancy Trimming}.  Parameters of \emph{Conure} are mainly from the bottom embedding layer, middle layers, and the task-specific prediction layers. Both embedding and middle layers are allowed to be pruned. Despite that, 
we empirically find that the performance will not be affected even we keep
parameters in the
the embedding layer fixed after training $T_1$. This is very likely because
 lower-level features are more task-agnostic, similarly as in \cite{yuan2020parameter}.
The middle layers of TCN consist of the temporal convolutional layers with bias terms, normalization and ReLu layers (see Figure~\ref{prelim} (d)). 
Given that the normalization layer and bias terms have very few parameters,
we can keep them fixed for simplicity after training $T_1$.\footnote{\scriptsize Note that 
 we place the normalization layer before the TCN layer, as shown in Figure~\ref{prelim} (d), otherwise, we strongly suggest optimizing it along with the TCN layer in the following tasks.}
Thereby, we conclude
that to endow the continual learning capacity to \emph{Conure}, one just needs to manipulate parameters of the hidden (convolutional) layers.  This property is desirable as it helps \emph{Conure} to reduce task-specific overhead in both computation and storage,  and makes the learning process and parameters more manageable. Besides, we find that such property is also applicable to other types of networks, such as self-attention based Transformer~\cite{kang2018self} (see Section~\ref{Adaptability}).

The pruning process of $T_1$ is illustrated in Figure~\ref{conure_para} (a). To facilitate discussion, we describe it by using a  convolutional layer.
Formally, let 
$\bm{Z}_{T_1}\in \mathbb{R}^{a\times b}$
be the  weight of a convolutional layer, where $a\times b$ is the weight shape\footnote{\scriptsize Note the original convolutional kernel has a 3D shape, here we simply reshape it to 2D for better discussion. This process applies to weights with any shape or dimension.}.
Assume  we need to prune away $Q_{T_1}$ (e.g., $Q_{T_1}=70\%$) parameters
on $T_1$. Before pruning, we 
rank all parameters  (from the smallest to the largest) by a score function $g(\bm{Z}_{T_1})$, where $g(\bm{Z}_{T_1}^k)= |\bm{Z}_{T_1}^k|$ in this paper.\footnote{\scriptsize `$||$' denotes the symbol of absolute value.} 
 Correspondingly, we obtain the threshold value $\delta$ with index $Q_{T_1}*h (\bm{Z}_{T_1})$,
 where $h (\bm{Z}_{T_1})$ is the number of parameters in $\bm{Z}_{T_1}$.  $\delta$ distinguishes the less important parameters from important ones. 
To realize pruning, we introduce a binary mask 
$\bm{G}_{T_1}\in \{0,1\}$ 
with the same shape as $\bm{Z}_{T_1}$, defined by
\begin{equation}
\label{pruneG}
\bm{G_{T_1}}^{k}=\left\{\begin{matrix}
1 & g(\bm{G}_{T_1}^{k})>\delta \\ 
0& g(\bm{G}_{T_1}^{k})<\delta
\end{matrix}\right.
\end{equation}
The effective weights after pruning becomes $\bm{Z}_{T_1}\odot\bm{G}_{T_1}$, where $\odot$ is element-wise product operator.
This is reflected in Figure~\ref{conure_para} (a), where white cells denote these trimmed redundant parameters, and their values are set to zero when performing convolution. Finally, these pruning masks $\bm{G}_{T_1}$ for all convolutional layers
are saved for the next training stage. 

\noindent\textbf{Retraining}. As shown in Figure~\ref{pruning}, in the beginning \emph{Conure} will experience a decline in performance by using the pruned structure, due to big changes in neural network structure.  To regain its original performance, \emph{Conure} performs
retraining on the pruned architecture as demonstrated in  Figure~\ref{conure_para} (b). Due to the existence of  $\bm{G}_{T_1}$, only important parameters are re-optimized, while pruned parameters ($\bm{Z}_{T_1}\odot(\bm{X}-\bm{G}_{T_1})$) whose values are set to zero keep unchanged because  no gradients are created for them\footnote{\scriptsize $\bm{X}= ones (\bm{G}) $ is a tensor with all elements one.}. As shown from (a) to  (b), parameters represented by the 
light red cells are modified to new ones with dark red colors,  from $\bm{Z}_{T_1}\odot\bm{G}_{T_1}$ to $\bm{\hat{Z}}_{T_1}\odot\bm{G}_{T_1}$.
We refer to  $\bm{\hat{Z}}_{T_1}\odot\bm{G}_{T_1}$  as condensed parameters of $T_1$, which keep fixed at this point onwards. After a period of retraining, the performance on $T_1$ is very likely to recover   as long as the pruning percentage is not too large. The updated parameters $\bm{\hat{Z}}_{T_1} $ are  saved to replace the original $\bm{Z}_{T_1}$ for the next stage. 

The pruning and retraining operations on $T_i$ ($i> 1$) will  be executed  only on task-specific parameters of $T_i$, where important parameters from  $T_{1}$ to $T_{i-1}$ are not allowed to be modified. 
For example, after training $T_2$, \emph{Conure} once again performs pruning and retraining to prepare it for $T_3$. As shown in Figure~\ref{conure_para} (d) and (e), only green cells from $T_2$ are pruned, while all red cells keep fixed. This allows \emph{Conure} to always focus on optimization of the task at hand.

\noindent\textbf{New task Training via knowledge retention}. 
At this phase, \emph{Conure} is required to
accomplish two goals:  i) achieving positive transfer on the new task $T_i$ by leveraging  condensed parameters  (i.e., dark color cells in Figure~\ref{conure_para}) from  $T_{1}$ to $T_{i-1}$; ii) overcoming forgetting these condensed parameters when learning  $T_i$. To this end,
 we only allow the redundant parameters of $T_i$ to be modified whereas condensed parameters from all past tasks are employed as prior knowledge and kept frozen only for forward propagation.
The weight used for learning $T_i$, i.e., $\bm{Z}_{T_i}$, is given as:
 \begin{equation}
 \small
 \label{Zstage4}
 \bm{Z}_{T_i}=\underbrace{\bm{\hat{Z}}_{T_{i-1}}\odot(\bm{X}-\sum_{j= 1}^{i-1} \bm{G}_{T_j}) }_{\mathsf{freed \ up \ parameters}}+\underbrace{\mathsf{stop\_gradient} (\bm{\hat{Z}}_{T_{i-1}}\odot \sum_{j=1}^{i-1} \bm{G}_{T_j})}_{\mathsf{past \ condensed \  parameters}}
 \end{equation}
where $\bm{\hat{Z}}_{T_{i-1}}$ is the weight of $T_{i-1}$ after retraining, $\bm{G}_{T_j}$ is the task-specific weight
mask generated by pruning, 
and  $\mathsf{stop\_gradient}$ is an operator that prevents the gradient from back propagation.  For example, by performing training on $T_2$, the white cells  are activated to light green, as shown in Figure~\ref{conure_para}~(c), while the dark red cells (condensed parameters) are kept unchanged. 
Following this way, \emph{Conure} could perform iterative redundancy pruning and parameter retraining for new coming tasks so as to add more tasks into the backbone network.  
This process can be repeated until all tasks are added or no free capacity is available.

\noindent \textbf{Overhead}. 
In contrast to the sequential training described in section~\ref{lst},
 \emph{Conure}  incurs additional storage overhead by maintaining the sparse mask $\bm{G}_{T_i}$.  However, as analyzed, if (after pruning) a parameter is useful  for $T_i$ , then it is used for all the following tasks $\{T_{i+1},...,T_{N}\}$, and meanwhile, it had actually been ignored for all the past tasks $\{T_1,...,T_{i-1}\}$.
 This means  the values corresponding to these parameters in the masks before and after $T_i$ will be set as zero. 
Thus, the total number of additional non-zero (i.e., one) parameters in these sparse masks of all tasks in $\textbf{T}$  is  upper-bound to the size of the convolution parameters in the backbone network.  Hence, \emph{Conure} is much more parameter-efficient than the individually trained network for each task.

 
\noindent \textbf{Inference}. 
Once given a selected taskID, we can obtain the inference network of  \emph{Conure}  which has the same structure as that developed for
 training for this task. Its only computation overhead is the masking operation which is implemented by multiplying convolution kernels with sparse tensors in an element-wise manner.

\section{Experiments}
We assess the sequentially learned user representations by \emph{Conure} on two tasks: personalized recommendations \& profile predictions. 
\label{EXPERIMENTS}
\subsection{Experimental Settings}
\label{expset}
\begin{table} 
	\centering
	\caption{\small Number of instances.  
		The number of distinct items  $|\mathcal{X}|$ in $T_1$ for TTL and ML  is $646K$ and $54K$ ($K=1000$), respectively.
		The number of  labels $|\mathcal{Y}|$ is $18K$, $ 8K$, $8$, $2$, $6$, respectively from $T_2$ to $T_6$ in TTL, and $26K$, $16K$, respectively from $T_2$ to $T_3$ in ML. $M=1000K$.
	}
	\small
	\label{datasets}
	\setlength{\tabcolsep}{2.34mm}
	\begin{threeparttable}				
		\begin{tabular}{c|c|c|c|c|c|c}
			\toprule
			\small Data &  \small $T_1$& \small $T_2$ &  \small $T_3$&  \small $T_4$&  \small $T_5$&  \small $T_6$\\
			\midrule
			TTL       &$1.47M$ &$2.70M$ &$0.27M$&$1.47M$&$1.47M$&$1.02M$\\ 	
			\midrule
			ML     & $0.74M$ & $3.06M$ & $ 0.82M$&-&-&-\\ 				
			\bottomrule
		\end{tabular}
	\end{threeparttable}
	\vspace{-0.1in}
\end{table}

\begin{table*} 
	\centering
	\caption{\small Accuracy comparison.  \#B is the number of backbone networks. 
		The left and right of `||' represent TTL and ML, respectively. 		\emph{Conure-} denotes \emph{Conure} that has not experienced the pruning operation after training on the current task. The worse and best results are marked by `$\triangledown$'  and `$\bigtriangleup$', respectively.
	}
	\small
	\label{overallcom}
	\setlength{\tabcolsep}{3.3mm}
	\begin{threeparttable}				
		\begin{tabular}{l|c|c|c|c|c|c|c||c|c|c|c}
			\toprule
			\small Model &  \small $T_1$& \small $T_2$ &  \small $T_3$&  \small $T_4$&  \small $T_5$&  \small $T_6$ &  \small \#\textbf{B} &\small $T_1$&  \small $T_2$&  \small $T_3$&  \small \#\textbf{B} \\
				\midrule
	DNN       &0.0104 &0.0154 &0.0231&0.7131&0.8908&0.6003&  \textbf{6} &0.0276&0.0175 &0.0313 &    \textbf{3}\\   \midrule
			\midrule
			SinMo       &0.0473 &0.0144&0.0161&0.7068&0.8998&0.5805$^\triangledown$&  \textbf{6} &0.0637&0.0160&0.0259$^\triangledown$ &    \textbf{3}\\ 
			\midrule
			SinMoAll    &0.0009$^\triangledown$ &0.0079 $^\triangledown$ &0.0124$^\triangledown$&0.5640$^\triangledown$&0.7314$^\triangledown$&0.6160&   \textbf{1}&0.0038$^\triangledown$&0.0145$^\triangledown$&0.0310 &    \textbf{1}\\ 	
			\midrule
			FineSmax     & 0.0473  &0.0160 & 0.0262&0.6798&0.8997&0.6070&    \textbf{1}&0.0637&0.0150&0.0262&    \textbf{1}\\ 				
			\midrule
			FineAll  & 0.0473  &0.0172& 0.0271&0.7160$^\bigtriangleup$&0.9053&0.6132&    \textbf{6}&0.0637&0.0189 &0.0325&   \textbf{3}\\ 				
			\midrule
			PeterRec     & 0.0473  & 0.0173 & 0.0275 &0.7137&0.9053&0.6156&    \textbf{1}&0.0637&0.0182&0.0308&   \textbf{1}\\ 
			\midrule
				MTL     & - & 0.0151 & 0.0172&0.7094&0.8979&0.6027&   \textbf{1}&-&0.0167&0.0276&   \textbf{1}\\ 				
			\midrule
			\midrule
			\emph{Conure-}     & 0.0473  & 0.0174 &0.0286&0.7139&0.9051&0.6180&  -&0.0637&0.0183&0.0347&   -\\ 
			\midrule
			\emph{Conure}     & 0.0480$^\bigtriangleup$  & 0.0177$^\bigtriangleup$& 0.0287$^\bigtriangleup$ &0.7146&0.9068$^\bigtriangleup$&0.6185$^\bigtriangleup$&   \textbf{1}&0.0656$^\bigtriangleup$&0.0197$^\bigtriangleup$&0.0353$^\bigtriangleup$&    \textbf{1}\\ 					
			\bottomrule
		\end{tabular}
	\end{threeparttable}
\end{table*}

{\textbf{Datasets.}} 
As for the first work in continual UR learning over tasks,
we find two public datasets to back up our key claim. They are
the Tencent TL dataset released by~PeterRec~\cite{yuan2020parameter}, referred to as TTL\footnote{\scriptsize \url{https://drive.google.com/file/d/1imhHUsivh6oMEtEW-RwVc4OsDqn-xOaP/view?usp=sharing}}, and the movielens\footnote{\scriptsize \url {https://grouplens.org/datasets/movielens/25m/}} dataset, referred to as ML. To be specific, TTL includes six different datasets connected by userID --- three for item recommendations and three for profile classifications. Each instance $(u,\bm{\mathrm{x}}^u)$ in the dataset of $T_1$  contains a userID and his recent 100 news \& video watching interactions  on the \textit{QQ Browser}  platform;  each instance  $(u, y)$  in $T_2$ contains a userID and one of his clicking (excluding thumbs-up) interactions 
on the \textit{Kandian}
platform;
each instance  in $T_3$ contains  a userID and one of his  thumb-up interactions  on \textit{Kandian},
where thumb-up  represents more satisfactory than clicks.
Each instance  in $T_4,\ T_5, \ T_6$ contains a userID and his/her age, gender, and life status categories, respectively. 
We apply  similar pre-processing for ML to mimic an expected CL setting, where  each instance in $T_1$  contains a userID and his recent 30 clicking  (excluding 4- and 5-star) interactions, each instance in  $T_2$ contains a userID and an  item that is rated higher than 4, and
each instance  in  $T_3$ contains a userID and one of his 5-star items.
A higher star means more satisfactory, the prediction of which is regarded as a harder task.
 Table \ref{datasets} summarizes the dataset statistics.

\noindent{\textbf{Evaluation Protocols.}} 
To evaluate  \emph{Conure}, we randomly split each dataset in $T_i$ into training (80\%), validation (5\%) and test  (15\%). We save parameters for each model only when they achieve the highest accuracy on the validation sets, and report results on  their test sets.
We use the popular top-$N$ metric MRR@$5$ (Mean Reciprocal Rank)~\cite{yuan2016lambdafm} to measure the recommendation tasks,
and the classification accuracy (denoted by Acc, where Acc = number of correct predictions$/$total number of instances) to measure the profile prediction tasks. 

\noindent{\textbf{Compared Methods.}} 
So far, there is no existing  baseline for  continual UR learning over different tasks. 
To back up our claim, 
We first present
a typical two-layer DNN network following~\cite{covington2016deep} for reference,  where for learning $T_i \ (i>1)$ the interaction sequence in $T_1$ is used as the outside features.
Note we have omitted baselines such as DeepFM~\cite{guo2017deepfm} and NFM~\cite{hexiagnan2017neuralf} since in ~\cite{yuan2020parameter}, authors showed that FineAll and PeterRec outperformed them.
Except DNN, all of them apply the same TCN network architecture, shared hyper-parameters and sequential learning pipelines (except MTL trained simultaneously) for strict and meaningful comparisons.
\begin{itemize}
	\item  \textbf{SinMo}: Trains a single model for every task from scratch and applies no transfer learning between tasks. SinMo uses the same network architectures  as \emph{Conure} in each training stage (see Figure \ref{prelim} (a) (b) and (c)) but is initialized randomly. 
	\item  \textbf{SinMoAll}: Applies a single backbone network for all tasks trained one by one without  preserving parameters learned from previous tasks, as described in Section~\ref{lst}. 
	\item  \textbf{FineSmax}: After training $T_1$, only the final softmax layer for $T_i$ ($i>1$) is finetuned, while all parameters from its backbone network are kept frozen \& shared throughout all tasks.
	\item  \textbf{FineAll}: After training $T_1$, all parameters for $T_i$ are finetuned. To avoid the forgetting issue in SinMoAll, it requires to maintain additional storage for parameters of each task.
	\item  \textbf{PeterRec}: Is a parameter-efficient transfer learning framework which
	 needs to maintain only a small number of separate parameters for the model patches and softmax layers, while all other parameters are shared after $T_1$, see~\cite{yuan2020parameter}.
	\item  \textbf{MTL}: Is a standard multi-task optimization via parameter sharing~\cite{caruana1997multitask} in the backbone network. Since not all users have training labels in each task, we perform MTL only using two objectives, one is $T_1$ and the other is $T_i$ ($i>1$). 
\end{itemize}

\noindent{\textbf{Hyper-parameter.}}
We assign the embedding \& hidden dimensions $f$ to 256 for all methods since further increasing 
 yields no obvious accuracy gains.
The learning rate is set to 0.001 for $T_1$ and 0.0001  for other tasks, similar to PeterRec. We use the Adam~\cite{kingma2014adam} optimizer in this paper. The regularization coefficient  is set to 0.02 for all tasks except $T_3$ on TTL,  where it is set to 0.05 for all models due to the overfitting problem.
All models use dilation $4\times \{1,2,4,8\}$ (16 layers)  for TTL  and  $6\times \{1,2,4,8\}$ (24 layers)
for ML. 
The batch size $b$ is set to 32 for $T_1$ and 512 for other tasks due to GPU memory consideration.
Following  PeterRec, we use the sampled softmax for $T_1$ with 20\% sampling ratio. 
The popularity-based negative sampling coefficient is set to 0.3 (a default choice in~\cite{yuan2016lambdafm}) for $T_2$  and $T_3$ for all models. 
\subsection{Performance Comparison and Insights}
\label{overallperform}
We  perform sequential learning on
the training sets of all tasks from $T_1$ to $T_i$ ($i=6 \ \& \ 3$ for TTL and ML, respectively ),  and  then evaluate  them on their test sets.
The pruning ratios of  \emph{Conure}  are 70\%, 80\%, 90\%, 80\%, 90\%, 90\%, respectively from $T_1$ to $T_6$  on  TTL (with 32\% free parameters left), and 70\%, 80\%, 70\%, respectively from $T_1$ to $T_3$ on  ML (with 39\% free parameters left).  We report results in Table~\ref{overallcom}.

\noindent \textbf{Catastrophic forgetting}: We observe that SinMoAll performs the worst among all tasks  except the last one (i.e., $T_6$ of TTL, and $T_3$ of ML). It is even much worse than SinMo which has no transfer learning between tasks. 
This is because SinMoAll uses one backbone network for all tasks, suffering from severe catastrophic forgetting for its past tasks --- i.e., after learning $T_i$, most parameters for $T_1$ to $T_{i-1}$ are largely modified, and therefore it cannot make accurate prediction for them anymore. Nevertheless, it yields relatively good results on $T_6$ as there is no forgetting for the last task. 
In contrast, \emph{Conure} clearly exceeds SinMoAll by overcoming forgetting although it also employs only one backbone network.

\noindent \textbf{One-time TL from $T_1$ to $T_i$} (e.g.,  $T_1\rightarrow T_2$, $T_1\rightarrow T_3$, or $T_1\rightarrow T_6$): SinMo shows worse results (after $T_1$) comparing to other baselines because of  no  transfer learning between tasks. By contrast, FineAll produces much better results, although the two models share  exactly the same network architecture and hyper-parameters. The main advantage of FineAll is that before training each $T_i$  ($i\geq2$), it has already obtained a well-initialized representation by training $T_1$.

 \emph{Conure} and PeterRec perform competitively with FineAll on many tasks, showing their capacities in doing positive transfer learning from $T_1$ to other tasks. But compared to FineAll, \emph{Conure}  and PeterRec are parameter very efficient since only one backbone network is applied for all tasks. In addition,
 FineAll largely surpasses FineSmax,  indicating that only optimizing the final prediction/softmax layer is not
expressive enough for learning a new task.

\noindent \textbf{Multiple TL from $T_1$ to $T_i$} (e.g., $T_1\rightarrow \mathbf{T_2}\mathbf{\rightarrow} \mathbf{T_3}, ... ,\rightarrow T_6$): Compared to FineAll and PeterRec,
\emph{Conure-}  yields around  4\% and 7\% accuracy gains on $T_3$ of TTL
and ML, respectively.
The  better results are mainly from the positive transfer from $T_2$ to $T_3$, which cannot be achieved by
any other model. To our best knowledge, so far \emph{Conure} is the only model which could
keep positive transfer learning amongst three or more tasks.
Another finding is that \emph{Conure-} does not obviously beat PeterRec and FineAll on $T_4$, $T_5$  and $T_6$. We believe that this is reasonable since there might be no further positive transfer from $T_2$, $T_3$ to   $T_4$, $T_5$, $T_6$ given that it has experienced one-time effective transfer from $T_1$ to $T_4$, $T_5$, $T_6$.\footnote{\scriptsize Note it is not easy to find an ideal publicly available dataset, where all tasks share expected similarities. But in practice, there indeed exist many  related tasks for both recommendation and profile prediction. For example, a recommender system may require to predict user various interactions, such as clicks, likes, comments, shares, follows and reposts, etc; similarly, a profile predictor may estimate user's profession, education and salary, which could potentially have some high correlations.} 
But  the good point  is that   \emph{Conure} does not become worse even when there is no effective positive transfer. The slightly improved result of
 \emph{Conure-} on $ T_6$ mainly comes from its robustness since parameters of irrelevant tasks may act as good regularization to resist overfitting.
By comparing  \emph{Conure-} and \emph{Conure}, we find that properly pruning with retraining usually brings a certain percentage of improvements for deep user representation models.

\noindent \textbf{Performance of other baselines}:  MTL  outperforms SinMo, showing the effects by applying multi-objective learning since the only difference between them is an additional $T_1$ loss in MTL. But it still performs worse than FineAll, PeterRec and    \emph{Conure}.
One key weakness of MTL is that it has to consider the accuracy for all (i.e., 2) objectives simultaneously, and thus might not be always optimal for each of them. Besides, MTL is unable to leverage all training data since some users of $T_1$ have no training instances on $T_i$ ($i\geq2$). Meanwhile, the standard DNN performs relatively well on some tasks but much worse on $T_1$ because it is unable to model the sequential patterns in user actions.  Another drawback is that such models (including DeepFM and NFM) have to be trained 
individually for each task, which are parameter-inefficient as well.

\begin{table} 
	\centering
		\small
	\caption{\small Impact of $T_2$ on $T_3$.  \emph{Conure\_no$T_2$} denotes  training \emph{Conure} on $T_3$	after $T_1$.  \emph{Conure\_no$T_2$} and \emph{Conure} both are  the \emph{Conure-} versions.
		TTL20\% and   \small ML20\% denote the 20/80 train/test split.
	}
	\small
	\label{t2impact}
	\setlength{\tabcolsep}{3.8mm}
	\begin{threeparttable}				
		\begin{tabular}{l|c|c|c|c}
			\toprule
			\small  &   \small TTL&  \small TTL20\%  &\small ML &  \small ML20\% \\
			\midrule
			\emph{Conure\_no$T_2$}    & 0.0277  & 0.0245& 0.0334&0.0295\\ 	
			\midrule
			\emph{Conure} & 0.0286& 0.0261&0.0347 &0.0309\\ 	
				\midrule
			\emph{Impro.} & 3.2\% &6.5\%&3.9\% &4.7\%\\ 							
			\bottomrule
		\end{tabular}
	\end{threeparttable}
\end{table}

\vspace{0.05in}
\begin{table} 
	\centering
	\caption{\small  Impact of task orders. Order1 is the original order as mentioned in Section~\ref{expset}. KC, KT and Life denotes the clicking dataset, the thumbs-up dataset and the life status dataset of \textit{Kandian}, respectively. Results on $T_1$ are omitted due to the same accuracy. The left and right of `||' are results of  \emph{Conure-}  and  \emph{Conure}, respectively. 
	}
	\small
	\label{torders}
	\setlength{\tabcolsep}{2.3mm}
	\begin{threeparttable}				
		\begin{tabular}{c|c|c|c||c|c|c}
			\toprule
			\small Orders &  \small KC& \small KT &  \small Life &  \small KC &  \small KT &  \small Life\\
			\midrule
			Order1       &0.0174 &0.0286 &0.6180&0.0177&0.0287&0.6185\\ 	
			\midrule
			Order2     & 0.0174 & 0.0289 & 0.6154&0.0177&0.0290&0.6152\\ 		
			\midrule
			Order3     & 0.0174 &0.0289 & 0.6145&0.0177&0.0287&0.6149\\ 				
			\bottomrule
		\end{tabular}
	\end{threeparttable}
		\vspace{-0.1in}
\end{table}

\begin{figure}
	\small
	\centering     
		\subfigure[\scriptsize TTL $T_6$ (one epoch: 1000*b) ]{\label{yahoo-alphazero}\includegraphics[width=0.23\textwidth]{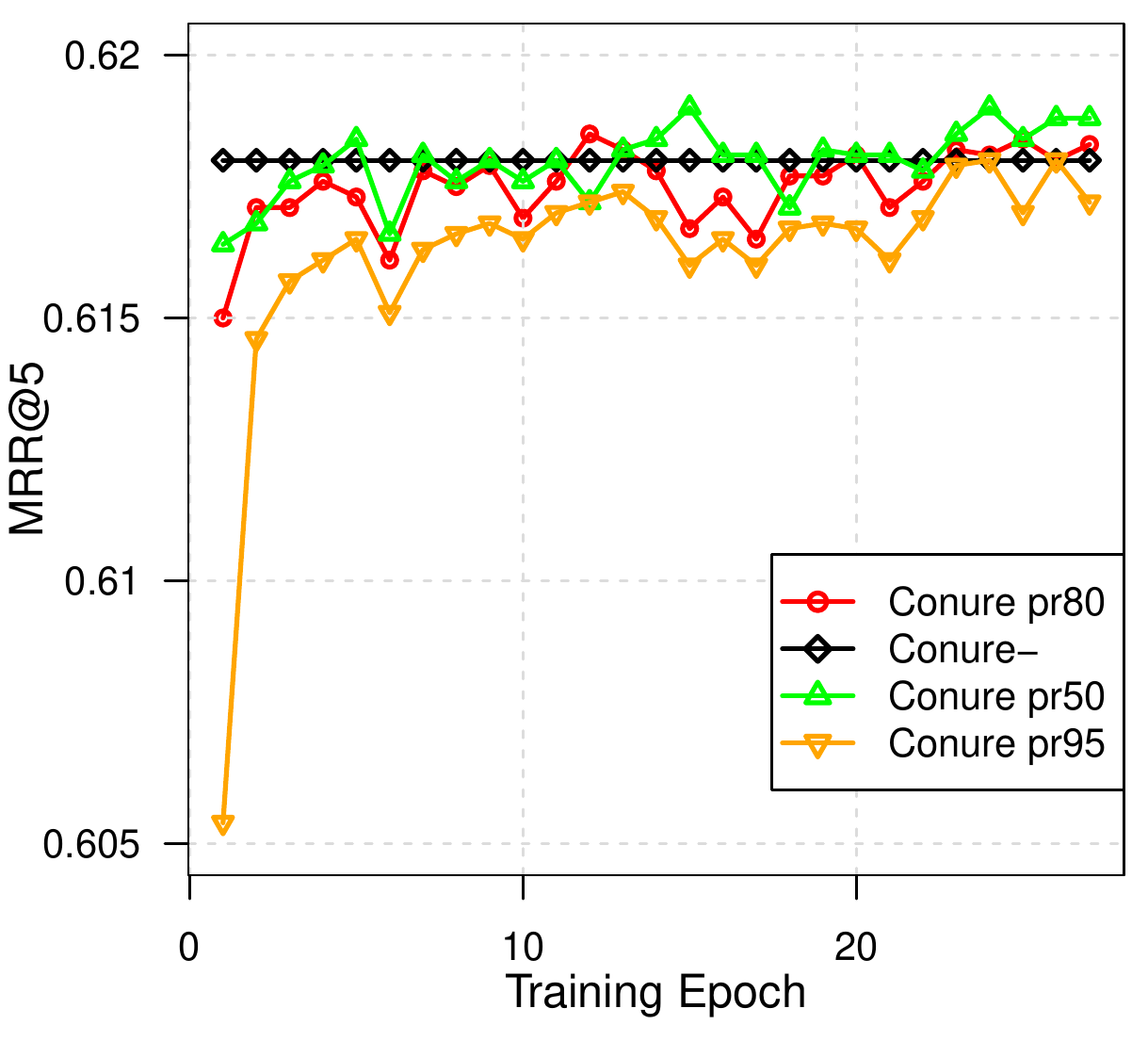}}
	\subfigure[\scriptsize  ML $T_2$ (one epoch: 2000*b) ]{\label{yahoo-alphazero}\includegraphics[width=0.23\textwidth]{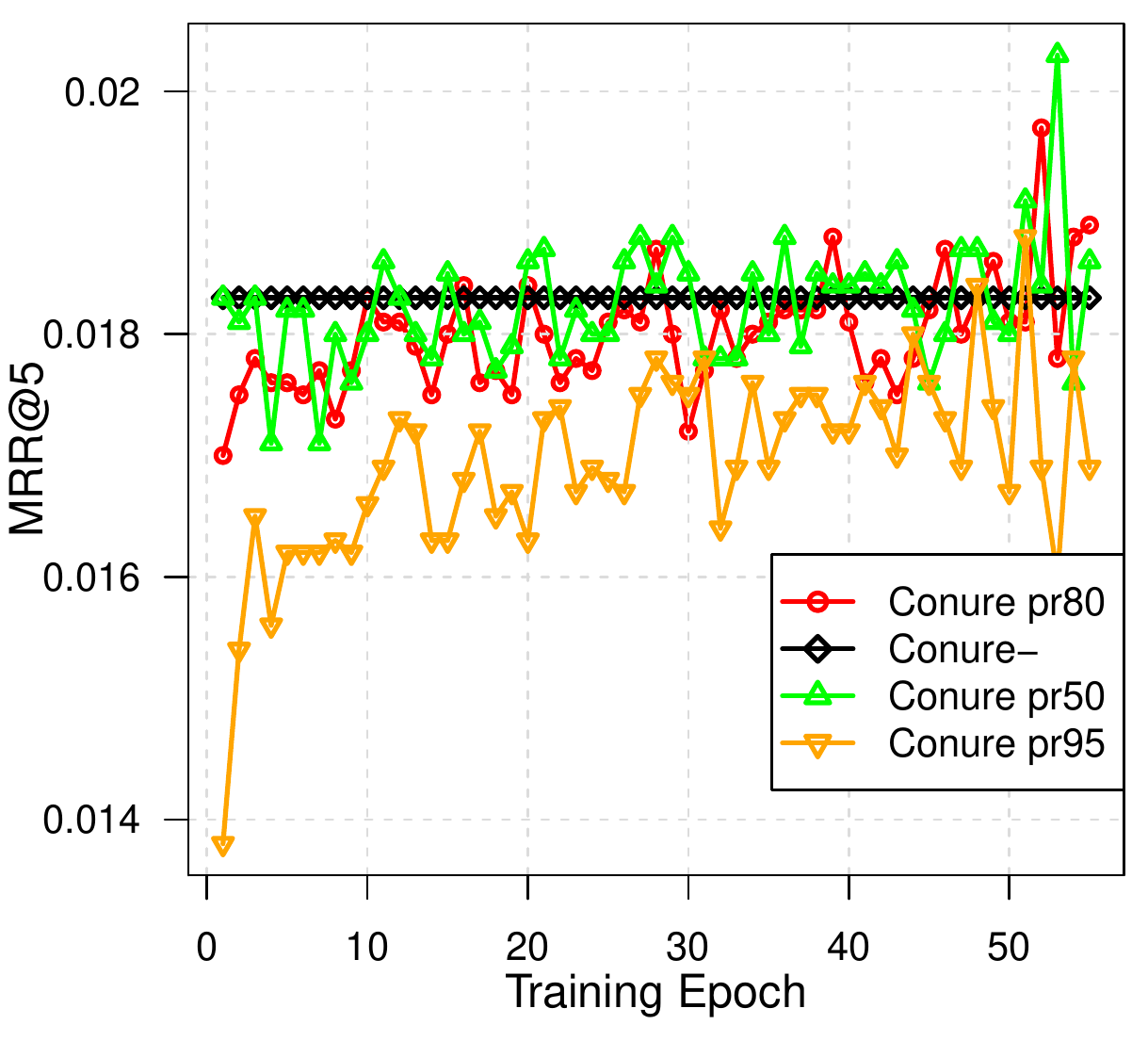}}
	\caption{\small Impact of pruning percentages.   The numbers (i.e., 50,  80, 95) denote  pruning ratios.
	}
	\label{pruneratio}
	\vspace{-0.2in}
\end{figure}
\subsection{Impact of $T_2$ for $T_3$}
We  perform more strict studies to verify the positive transfer from $T_2$ to $T_3$, since it is the unique ability of \emph{Conure} distinguishing from all other models. To do so, we evaluate \emph{Conure} on $T_3$ without training $T_2$ in advance. To clearly see the transfer effect, we also report results with a new split with 20\% data for training and the remaining for testing since TL may not be necessary if there is  enough task-specific data.  As shown in Table~\ref{t2impact}, \emph{Conure} clearly  outperforms \emph{Conure\_no$T_2$} on both TTL and ML (with statistical significance). Particularly, \emph{Conure} obtains 6.5\% accuracy gain  on TTL20\% by learning $T_2$ ahead.  Such findings well back up our claim and motivation regarding the advantage of  \emph{Conure} --- i.e., it is  particularly expert at the sequential tasks learning once they have a certain relatedness.

\subsection{Impact of Task Orders }
In this subsection, we are interested in investigating whether  \emph{Conure} is sensitive to the task orders. It is worth noting that the first task should not be changed since its  responsibility is to obtain the base user representation.
To be specific, we compare  \emph{Conure} with another two orders on TTL, namely $T_1 \rightarrow T_2 \rightarrow T_6 \rightarrow T_3$  (denoted as Order2) and  $T_1 \rightarrow T_6 \rightarrow T_2 \rightarrow T_3$ (denoted as Order3). As shown in Table~\ref{torders},   \emph{Conure} is in general not sensitive to task orders. Interestingly, \emph{Conure} performs better on Life when it is trained lastly (i.e., Order1).  One  reason may be that parameters of previous tasks could also work as good regularization terms, which 
increase model robustness to noisy labels and overfitting. Its accuracy on Life with Order2 \&  3 are
almost the same as PeterRec in Table~\ref{overallcom}, which is further evidence for this argument. Likewise,  \emph{Conure-} performs slightly better on KT with order2 \& 3, because KT is trained lastly.

\begin{table} 
	\centering
	\caption{\small Pruning and retraining both the embedding \& convolutional layers. The left \& right of `||' are tasks on TTL  \& ML.   }
	\small
	\label{pruneembs}
	\setlength{\tabcolsep}{2.35 mm}
	\begin{threeparttable}				
		\begin{tabular}{c|c|c|c||c|c|c}
			\toprule
			\small Models &  \small $T_1$& \small $T_2$ & \small $T_3$ &   \small $T_1$ &  \small $T_2$ &  \small $T_3$\\
			\midrule
			\emph{Conure-}       &0.0473 &0.0175 &0.0290&0.0637&0.0191&0.0341\\ 	
			\midrule
			\emph{Conure}     & 0.0474 & 0.0177  &0.0295 &0.0645&0.0196&0.0347\\ 					
			\bottomrule
		\end{tabular}
	\end{threeparttable}
\end{table}

\begin{table} 
	\centering
	\caption{\small Results by specifying \emph{Conure} with Transformer as the backbone network. The left and right of `||' represent tasks on TTL  and ML, respectively. `Mo' , `FA', `C-', `C',  denotes Models, FineAll, Conure- and Conure, respectively. 
	}
	\small
	\label{transformer}
	\setlength{\tabcolsep}{1.8 mm}
	\begin{threeparttable}				
		\begin{tabular}{l|c|c|c|c||c|c|c|c}
			\toprule
			\small Mo &  \small $T_1$& \small $T_2$ & \small $T_3$ & \small \#\textbf{B}  &   \small $T_1$ &  \small $T_2$ &  \small $T_3$ &  \small \#\textbf{B}\\
				\midrule
			\emph{FA}       &0.0510 &0.0161 &0.0243& \textbf{3}& 0.0654&0.0193&0.0321 &\textbf{3}\\ 	
			\midrule
			\emph{C-}       &0.0510 &0.0177 &0.0288 & - &0.0654&0.0198&0.0345 & - \\ 	
			\midrule
			\emph{C}     & 0.0513 & 0.0179  &0.0289 & \textbf{1} &0.0662&0.0200&0.0357 & \textbf{1} \\ 					
			\bottomrule
		\end{tabular}
	\end{threeparttable}
\end{table}

\subsection{Impact of Weight Pruning }

In this subsection, we examine the impact of pruning. We plot the retraining processes of $T_6$ (on TTL) and $T_2$ (on ML) in Figure~\ref{pruneratio}. As shown,  a few epochs of retraining after pruning  can recover the performance of \emph{Conure-}. In particular, \emph{Conure}  is able to outperform \emph{Conure-} even pruning away over 50\% redundant parameters. For example, \emph{Conure} improves MRR@5 of \emph{Conure-} from 0.0183 to 0.0203 when pruning 50\% parameters on $T_2$. In addition, we also notice that pruning too much percentage (e.g., 95\% ) of parameters  could lead to worse performance or slower convergence, as shown on (b). In practice, we suggest tuning the pruning ratios from 50\% to 80\%.

Though \emph{Conure} performs very well by performing continual learning on only middle layers, we hope to verify its applicability to the embedding layer. To this end, we prune and retrain
both the embedding  and convolutional layers. 
The pruning ratios for hidden layers remain the same as in Section~\ref{overallperform}, while
for the embedding layer they are 30\%, 80\%, 80\% for $T_1$,  $T_2$ and  $T_3$, respectively. 
As shown in Table~\ref{pruneembs}, we observe that  pruning and retraining additional parameters of the embedding layers reach similar results as in Table~\ref{overallcom}. An advantage by pruning the embedding layer is that more free capacity can be released to promote the future task learning.
\subsection{Adaptability}
\label{Adaptability}
Here we investigate whether the framework can be applied to other types of backbone networks. Inspired by the huge success of self-attention or Transformer in recent literature~\cite{devlin2018bert,vaswani2017attention}, we specify  \emph{Conure} with the Transformer architecture as the encoder.
We choose one attention head and two self-attention residual blocks due to its good performance in the validation set. Other hyper-parameters and setups are kept  exactly the same as in Section~\ref{expset}.  We prune and retrain only the linear transformation layers in the residual block (including weights from both the self-attention and feed-forward blocks). We report the results in Table~\ref{transformer}.
As shown, we basically achieve similar conclusions as before. Specifically, (i) compared with FineAll,
 \emph{Conure} obtains obvious improvement on $T_3$ on both TTL and ML, since FineAll could only enable one-time transfer learning, e.g., from $T_1$ to $T_3$, but \emph{Conure} could keep continual transfer learning from $T_1, T_2$ to $T_3$. (ii) \emph{Conure} requires only one backbone network for three tasks whereas FineAll requires three to avoid forgetting.  In addition, we also find that in contrast to TCN, 
 \emph{Conure} with Transformer as the backbone network usually yields some better results (see Table~\ref{overallcom}). But it is also worth noting Transformer requires quadratic time complexity to compute self-attention, whereas TCN has only linear complexity, which is much faster than Transformer when handling long-range interaction sequences.
\section{Conclusions and Impacts}
In this paper, we have confirmed two valuable facts: i) better user representations could be learned in a sequential manner by acquiring new capacities and remembering old ones; ii) continually learned user representations can be used to solve
 various user-related tasks, such as personalized recommender systems and profile predictions. 
We proposed \emph{Conure} --- the first task-level lifelong user representation model, which is 
conceptually very simple, easy to implement, 
and requires no \textit{very} specialized network structures. Besides, 
\emph{Conure} has achieved comparable or better performance in contrast to the classic learning paradigms (including single-task, multi-task and transfer learning) with minimal storage overhead.
We believe \emph{Conure} has made a valuable contribution in exploring the lifelong  learning paradigm for user representations, approaching the goal
 of one person, one model, one world.

For future work,  there is still much room for improvement of
\emph{Conure} towards a more intelligent lifelong learner.
First,  while \emph{Conure} is able to achieve positive transfer for new tasks, it could not in turn transfer the newly learned knowledge to improve old tasks.
 Second, while \emph{Conure}  can easily handle sequential  learning for over six tasks, it is yet not a never-ending learner  since it cannot automatically grow its architecture. Third, it is unknown whether the performance of \emph{Conure} will be affected if there are contradictory tasks
requiring optimization in opposite directions. We hope \emph{Conure} would inspire new research work to meet these challenges. 
We also expect some high-quality real-world benchmark datasets could be released so as to facilitate research in this difficult area.

\bibliographystyle{ACM-Reference-Format}
\bibliography{bibliography}

\appendix

\end{document}